\documentclass{PoS}

\title{Study of the performance of an array of Cherenkov telescopes by means of multi-objective evolutionary optimisation}

\ShortTitle{IACT Evolutionary Optimisation}

\author{\speaker{Bruno Fontes Souto}\footnote{Now at Instituto Federal de Educa\c{c}c\~{a}o, Ci\^{e}ncia e Tecnologia of Rio de Janeiro}\\
        Centro Brasileiro de Pesquisas F\'{i}iscas (CBPF), Rua Doutor Xavier Sigaud 150, 22290-180 URCA, Rio de Janeiro (RJ), Brazil.\\
        E-mail: \email{bruno.fontes.souto@gmail.com}}

\author{Ulisses Barres de Almeida\\
        Centro Brasileiro de Pesquisas F\'{i}iscas (CBPF), Rua Doutor Xavier Sigaud 150, 22290-180 URCA, Rio de Janeiro (RJ), Brazil.\\
        E-mail: \email{ulisses@cbpf.br}}

\abstract{This paper is concerned with the performance optimisation of an stereoscopic array of imaging atmospheric Cherenkov telescopes (IACTs) as a function of their positioning on the ground. In this first work we are concerned primarily with the study of the optimisation method and its test on toy arrays of few (3-6) telescopes. The ideas presented here were developed to investigate alternative ways of studying IACT array geometries. The proposal is an attempt to cover more exhaustively and systematically the parameter space involved in the design of a stereoscopic IACT array, aiming to develop a support tool for  directing the computationally expensive Monte Carlo simulations commonly used in the field. The methodology presented here involves a modelling step (in our case a simplified, heuristic IACT array model) and the implementation of an evolutionary algorithm for the geometric optimisation. In this initial work, the heuristic model and the optimisation algorithm are presented, but no detailed Monte Carlo validation is presented yet. The techniques used here may have potential applications in other optimization problems in the field of Gamma Ray Astronomy.}

\FullConference{36th International Cosmic Ray Conference -ICRC2019-\\
		July 24th - August 1st, 2019\\
		Madison, WI, U.S.A.}

\begin{document}

\section{Introduction}

The observation of gamma rays from space constitutes a privileged window into the astrophysical sites of particle acceleration and for the identification of the sources of cosmic rays. Upon interaction with the Earth's atmosphere, gamma rays give rise to a cascade of secondary particles that propagate towards the ground in what is called an {\it atmospheric shower}. Gamma ray-initiated (or lepton-initiated) showers -- which differ from the hadron-initiated ones by means of the physical processes governing their atmospheric development  -- are composed of secondary gamma-rays and electron-positron pairs which cascade down in a multiplicative process sustained by pair production. The shower is maintained for several km, until the secondaries' energy decreases below the threshold for ionisation of the molecules in the atmosphere, and secondary gamma-rays cease to be produced via bremsstrahlung. By detecting the Cherenkov photons produced as the result of the propagation of the superluminal particles of the shower, the {\it imaging atmospheric Cherenkov} (IAC) technique is able to reconstruct the energy and the arrival direction of the primary gamma-rays impinging in the atmosphere.

The present conference happens exactly 30 years since the IAC technique was first used with success to detect a very-high energy gamma-ray cosmic source, the Crab Nebula -- a feat performed by the pioneer Whipple Telescope in Arizona~\cite{Weekes1989}. Since then, well over a hundred sources have been discovered through the use of arrays of imaging atmospheric Cherenkov telescopes (IACTs)~\cite{Hillas2013}. Optimising the performance of these IACT arrays means increasing the rate of event detection and improving the reconstruction of the primary gamma-ray. Traditionally, this is done by means of Monte Carlo simulations, which are used to optimise both individual telescope parameters and the stereoscopic array geometry. Array arrangements of a few IACTs constitute in regular, symmetrical geometries, where the spacing between the telescopes represents a trade-off between an optimised stereo view of the showers, which favour good reconstruction of the events, and a large effective area for maximal sensitivity.

The high computational effort demanded from such detailed Monte Carlo simulations means that the optimisation of significantly larger arrays is a challenging task, and the complexity of the parameter space to be investigated considerably larger. For a thorough investigation of the problem, state-of-the-art softwares such as CORSIKA~\cite{Heck1998} and {\it sim-telarray}~\cite{Bernlohr2013}, which calculate the detailed shower development and take into account the full details of the shower detection process by the IACTs, are used. The strategy proposed in this paper to try to circumvent the computational cost involved in the full-scale simulations consists in translating the optimisation objectives into a heuristic model of the array, and write appropriate cost-functions. These functions can then be used in an optimisation algorithm to systematically investigate a larger parametric space of the array configuration than feasible with a direct Monte-Carlo approach. In our case, and to simplify this first investigation into the method, we are not concerned with optimisation of individual telescope properties, but only the array geometry, expressed in terms of the relative arrangement (positions) of the telescopes on the ground.

As an alternative to this approach, some authors have historically considered the concept of {\it array cells}. In this strategy, a "unitary cell" composed of a few telescopes which can simultaneously trigger a shower event is optimised (i.e., a squared arrangement of four Cherenkov telescopes), and this "cell" is then replicated in a regular grid to extend the array area, the full array performance scaling with the number $N$ of juxtaposed cells~\cite{Aharonian1997}. This simplistic approach has the drawback of being restricted to a single regular geometric arrangement of the telescopes throughout the whole array. Our more flexible is desirable in comparison with these previous exercises for being able to test (and accommodate) variations of this {\it local geometry} throughout the array, thus profiting of the large number of telescopes to have a more comprehensive response of the observatory to different, conflicting aspects of the shower reconstruction. 

The existence of conflicting objectives in the problem of IACT array optimisation motivates the use of Evolutionary Optimisation techniques. The idea is to create an efficient and easy-to-implement algorithm that generates near-optimal solutions for the array geometry. Being based on heuristics, the algorithm is not expected to provide fully optimised solutions, and its intent is not to replace the Monte Carlo approach for the final array optimisation. Rather, it aims to serve as a tool for efficiently investigating the complete parameter space of array geometries, and thus to open new hypothesis of array configuration to be detailedly investigated by full-scale simulations. The advantage of such heuristic approach based on evolutionary optimisation is its much lower computational cost. In other fields of astronomy, such as with radio interferometric arrays, similar approaches have been used for the array design, which resulted in the discovery of non-intuitive geometric solutions that outperformed the more usual, regular grid solutions~\cite{Cohanin2004}, similarly to what we aim to achieve here.

\section{Multi-objective optimisation and evolutionary algorithms}

Evolutionary algorithms are suitable approaches for solving multi-objective problems, such as the one in question here. They are based on the manipulation of a set of candidate solutions in order to obtain a progressive improvement through the sequencial iterations of the algorithms, in analogy with the successive generations of living beings~\cite{Deb2001}. The desirable characteristics of the evolutionary methods, that justify our choice for this optimisation method are: (i) Possibility of concurrently manipulating a set of solutions at each iteration, a requirement resulting from the multi-objective character of the problem; (ii) Availability of strategies to encourage diversity of solutions; (iii) Convergence of solutions with good processing speed; (iv) Non-dependence of hypotheses on the search space; (v) Flexibility to address different formulations of the problem.

When dealing with multi-objective optimisation, it is often the case that the objective functions conflict with each other, that is, there is no single solution that optimises all of them simultaneously. In this regard, one must incorporate an alternative notion of optimal solution of the problem. In such situations, an optimal solution can be defined as the set of parameters such that there is no other solution that improves the fitness of one goal without worsening the fitness of a another goal as consequence. In other words, the optimal solutions provided by the algorithm are those belonging to a so-called Pareto front~\cite{Deb2001}.

As a matter of fact, there may be an infinite number of such Pareto-optimal solutions to a given problem. If additional information on the relative importance among objectives is unknown, all Pareto-optimal solutions are equivalente, and a judgement of best solution among the set cannot be made. On the other hand, such diversity of solutions provided by the evolutionary algorithms can be seen as an asset of the approach. It provides, in fact, a set of responses to the optimisation problem that cover all the possible trade-offs among the conflicting objective functions, instead of an optimal solution for each independent objective, or a single solution based on a pre-defined choice of weights for each conflicting term of the problem. It remains thus a choice of the decision maker to evaluate all solutions and choose the one that suits best the purposes of the design for deeper investigation~\cite{Coello2007}.

\section{Formulation of the problem}

The general formulation of a multi-objective optimisation problem can be written as

%\begin{subequations}
\label{eq:moop}
\begin{eqnarray}
            min  f_i(\textbf{v})&          \:  \: \:  \: \: \: \: \:     i&=1,...,M \\
             g_j(\textbf{v})  \geq 0&        \: \:  \: \: \:   \: \:            j&=1,..., N_g \\
             h_k(\textbf{v})  = 0&           \: \: \:  \: \: \: \:         k&=1,...,N_h \\
            v_n^L \leq  v_i  \leq  v_n^H&    \: \: \: \: \:  \:   \:         n&=1,...,N_v 
\end{eqnarray}
%\end{subequations}

\noindent where $g_j$ and $h_k$ are constraints which apply to the optimisation functions of the problem, $f_i$, and $\textbf{v}$ is the vector of variables of the problem, subject to $N_v$ constraints.

In our particular problem of optimising an array of IACTs by means of the relative positioning of its $N$ telescopes on the ground, there are no specific constraints to the optimisation functions, so we can neglect  $g_j$ and $h_k$, and $\textbf{v}$ are the two dimensional spatial variables. Due to the finite size of a single telescope's effective area, there are some constraints to be considered on $\textbf{v}$, which determine the size of the optimisation region. Thus, the location of each telescope $i$ is given by the ordered pair $ (x_i, y_i) \in \mathbb {R} ^ 2 $, $ i \in \{1, ..., N \} $ subject to $ x^L < x_i < x^H $ and $ y^L < y_i < y^H $.

In formulating our problem, we consider four optimisation parameters -- effective area, mean trigger multiplicity, angular resolution and energy resolution --  that aim to describe the overall performance of an array of Cherenkov telescopes, and use them to derive the set of heuristic objective functions $f_i$ that will be later implemented in the optimisation algorithm. It is important to stress that, in following an heuristic approach to the derivation of the fitness functions, we are concerned with representing only a general performance description for each optimisation parameter.

\quad

I) The Effective Area $A_{eff}$ is given by the integral of the probability $P$ of detection of an event by $N$ telescopes over the shower impact area $A$, \emph{i.e.},

\begin{equation}
\label{eq:areaefarray}
A_{eff}  = \int \! P \, \mathrm{d}A. 
\end{equation}

Note that $P$ depends on: the relative positions of all $N$ telescopes; the choice of the minimum trigger multiplicity to validate a stereoscopic detection; and the trigger probability function of each telescope (in turn a function of the distance between the shower core and the telescope -- the shower impact parameter). In this work, we consider a minimum trigger multiplicity of three telescopes to validate a stereoscopic detection. 

II) The Mean Trigger Multiplicity $<k>$ is defined as the weighted mean

\begin{equation}
\label{eq:multimedia}
<k> =  \frac{\sum_{k = 3}^{N} A_k \cdot k}{A_{eff}},     
\end{equation} 

\noindent where $k$ corresponds to the number of telescopes that can be triggered by an event in a stereoscopic detection, $A_k = \int \! P_k \, \mathrm{d}A$. 
%\simeq \sum\limits_{i=1} P_i^{k} A_i $ and $P_i^{k}$ are the the probabilities of detection by only $ k \in \{3, 4, ..., N \} $ telescopes in each area element $ A_i $ on a discrete region. 
$P_k$ are the probabilities of detection by only $ k \in \{3, 4, ..., N \} $ telescopes and $A_{eff} = A_3 + A_4 + ... + A_N$. A high average trigger multiplicity indicates enhanced quality in the reconstruction of primary gammas over the hadronic background and also affects angular and energy resolutions, being an element to be maximised in the optimisation problem~\cite{Maier2017, Konopelko1999, Bernlohr2013}.

III) We adopt a definition of angular resolution as the circle around a simulated source containing 68\% of the reconstructed arrival directions~\cite{Maier2017}. Thus, motivated by~\cite{Bernlohr2013}, we define $\theta_{\mu \nu} \in [0 , 90]$ as the mean stereo angle of observation of events by any pair of telescopes $\mu$ and $\nu$, and we ascribe weights depending on the Hillas Parameters Size (S) and Length (l) to define\footnote{The Cherenkov radiation emitted by the shower imprints an image in the telescope camera in the form of an ellipse, whose shape and orientation can be characterised by a set of parameters first formulated by Hillas~\cite{Hillas1985}. Here we are interested in two of these parameters: \textit{Size}, which is the measure of the total number of photoelectrons recorded by the camera and \textit{Length}, which is the root mean square of the distribution of the photoelectrons in the camera along the major axis of an ellipse fitted to the image.}

\begin{equation}
\label{eq:xi}
\Theta = sin(\theta_{\mu \nu}) \cdot (S_{\mu}^{-1} + S_{\nu}^{-1})^{-1} \cdot (l_{\mu}^{-1} + l_{\nu}^{-1})^{-1},
\end{equation}

\noindent in such a way that the function $<\Theta>$ used in place of the angular resolution is then given by the weighted mean of $\Theta$ over the $N$ telescopes and the shower impact area, that is,

\begin{equation}
\label{eq:xi_metrica}
<\Theta> = \frac{ \int \! \Theta P \mathrm{d}A } {\int \! P \mathrm{d}A }. % \simeq  \frac{\sum_i P_i \Theta_i} {\sum_i P_i},
\end{equation}

According to the formulation above, the angular resolution should increase in direct proportionality to $<\Theta>$.

IV) Finally, the energy resolution can be defined as the width of the distribution generated by the relative error between the reconstructed and simulated energies for a number of gamma-ray detected events (\emph{e.g.},~\cite{Aleksic2016}). It can be formulate din terms of the impact parameter of the shower, as observed by a single telescope\footnote{The impact parameter of the observation is the distance between the telescope and the position of the shower core on the ground.}. The fluctuations $L$ of the number of photoelectrons detected by a telescope for a series of gamma-ray events are a good proxy for the energy resolution (\emph{e.g.}~\cite{Aleksic2016}). Hence, we take as our parameterisation $<L>$, the mean value of these fluctuations for the $N$ telescopes that detect a given gamma-ray event in stereoscopic mode, integrated over the shower impact area

\begin{equation}
\label{eq:chi}
<L> = \frac{ \int \! L P \mathrm{d}A } {\int \! P \mathrm{d}A }. 
%\simeq  \frac{\sum_i P_i L_i} {\sum_i P_i},      
\end{equation}   

Using these four functions and the definition of the spatial variable constraints, we can now write our optimisation problem as

\label{eq:moop_2}
\begin{eqnarray}
            && max  \{ A_{ef}, <k>, <\Theta> \}         \\  
            && min  <L>                                 \\
            && x^L \leq  x_i  \leq  x^H \qquad i=1,...,N \\
            && y^L \leq  y_i  \leq  y^H \qquad i=1,...,N    
\end{eqnarray}

\noindent where $ \bf{v} = [\bf{x}^T \ \bf{y}^T]^T  \in \mathbb{R}^{2N}$ is a columm vector with the coordinates that determine the position of each telescope on the ground.

\section{The optimisation algorithm}

Now we present the algorithm developed to solve the many-objective optimisation problem presented in the previous section. In order to deal with the relatively large number of (four) objective functions, we use the evolutionary algorithm EliteNSGA-III, described in~\cite{Deb2014, Ibrahim2016}. There are three particular characteristics of the EliteNSGA-III that motivated our choice for using it, namely,

\begin{itemize} \itemsep -0.5ex 
\item[(i)] Aptitude to treat many (more than three) objectives, in the context of many-objective optimisation;
\item[(ii)] Lowest possible dependence on external evolutionary parameters - only recombination $n_c$ and mutation $n_m$ indexes were used;
\item[(iii)] The maintenance of the Pareto Criterion for non-dominated solutions.
% - which is an issue in problems with more than four objectives
\end{itemize}

Next we present a more detailed description of the algorithm, from its initialisation to its stopping criteria.

\begin{itemize} \itemsep -0.5ex 
\item[(i)] \textbf{Initiatilisation:} A solution for the array is an individual of $2N$ real numbers corresponding to the coordinates of the $N$ telescopes. The initial population is generated randomly and uniformly, with a number of individuals equal to the least even number equal or greater than $H$. The first elite, with the same size as the initial population, is initialised with the worst values for each objective and the telescope's positions at infinity;
\item[(ii)] \textbf{Reference points:} Firstly, the objective functions are normalised in the interval $[0,1]$. Then, $H$ equally spaced reference points are added into the objective space. We define $p$ as the number of spacings between the reference points along any coordinate axis;
\item[(iii)] \textbf{Creation of the offspring solutions:}  At each iteration, the algorithm works with two generations, $P$ and $Q$. Let $P_t$ be the population created by the candidates to solve the problem at iteration $t$ of the algorithm. The descending population $Q_i$ is generated from $P_t$ by simulated binary recombination~\cite{Agrawal1995} and polynomial mutation~\cite{Deb2001}. Finally, these two populations compete with each other to generate a new population $P_{t+1}$.
\item[(iv)] \textbf{Genetic operators:}  The recombination occurs among individuals of $P_t$ and the elite, where each individual of the population or elite is taken with 50\% probability of participation. The probability of a gene in each individual being mutated is taken to be equal to the inverse of the number of variables of $\textbf{v}$. 
\item[(v)] \textbf{Sorting \& Diversity:}  Using the non-dominance criterion, the solutions in $R = P \cup Q$ are ranked in non-dominated fronts, $F_{i}$. Given an iteration of the algorithm, the population $P_{t+1}$ will be generated by the non-dominated solutions of the fronts in $R_t$, receiving one front at a time. Since $P$ has a fixed size equal to $N$ and $P \cup Q$ has $2N$ solutions, not all fronts will be accommodated. When the last front, $F_{l}$, is admitted, the reference points are used to select which solutions will finally be accommodated. The fronts $\{F_i \ |\  i > l\}$ will be discarded. 
\item[(vii)] \textbf{Elite:} Each reference point has at most one member of the elite associated with it. After the population $P_{t+1}$ is generated, for each reference point, we find which of the solutions associated with it have the lowest norm. If this is less than the norm of the elite member already associated with the that point in the previous generation, then it will become part of the elite, replacing the previous member.

\end{itemize}

The algorithm stops after reaching a given number of generations. The ideal points of each of the $M$ goals of the optimisation were updated at each generation, to avoid the need to solve $M$ scalar optimisation problems before applying the multi-objective method.

\section{Example implementation and discussion}

\begin{figure}[t!]
	\center{
	\includegraphics[width=1\textwidth]{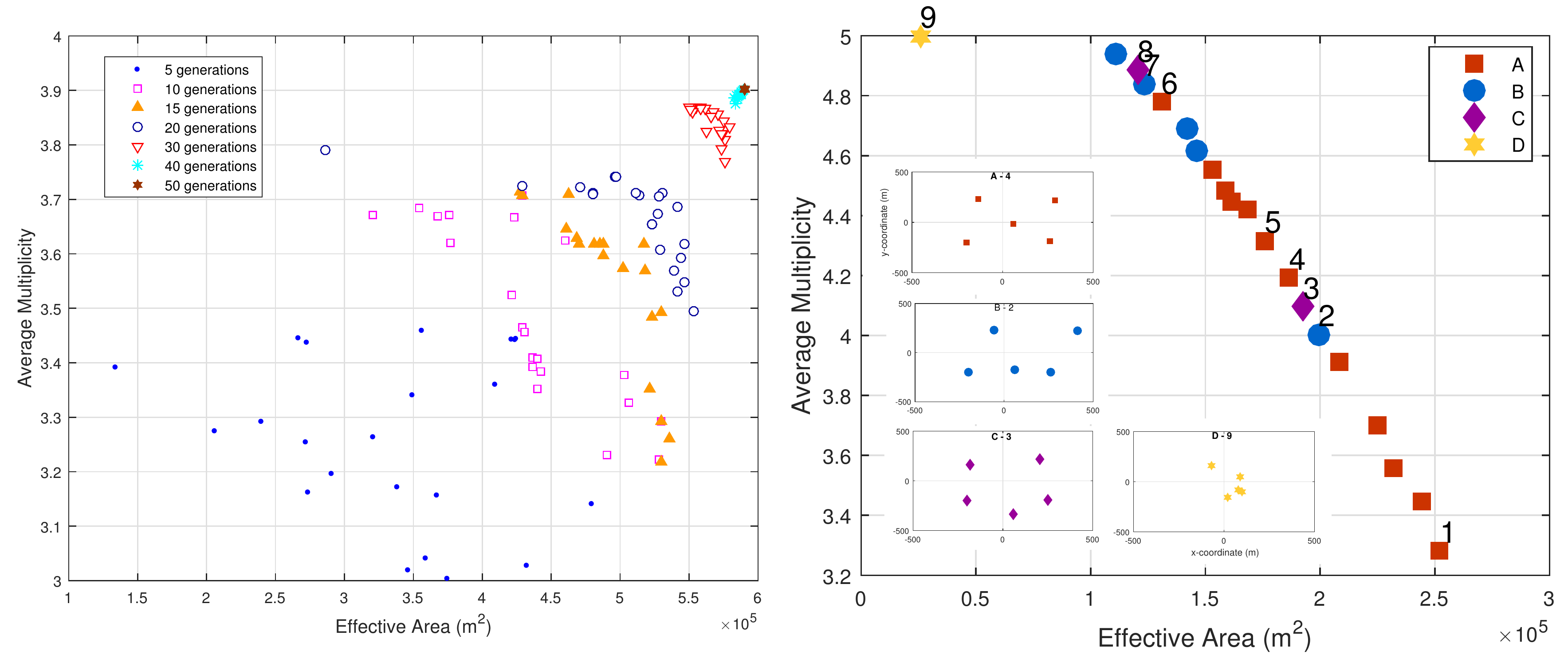}}
	\caption{(Left) Effective Area and Trigger Multiplicity as a function of the number of generations for a 4 telescope arrangement. Notice the convergence of the populations to a final single optimal geometric arrangement. (Right) Pareto Front for the effective area and average trigger multiplicity for 5 telescope arrangements. Similar geometrical configurations arisen from the trade-off between the two conflating objectives are group by different points and displayed in the inlet figures.\label{fig:figura}}
\end{figure}

In this section we show a couple of simple examples to illustrate the work of the algorithm. For simplicity, and given the limited space available in this paper, we present examples considering the evolutionary optimisation with only two of the four objective functions presented, namely, the effective area $A_{eff}$ and the average trigger multiplicity $<k>$. 

Figure 1 (left) exemplifies how the two objective functions $A_{eff}$ and $<k>$ for an entire population of 4-telescope arrays, satisfactorily converge to a single optimised arrangement, after 50 generations. For 5 telescopes (Figure 1, right), the algorithm has a similar convergence rate, but here the convergence produces a family of solutions (a Pareto Front), instead of a single solution. As we move along the Pareto Front, towards its extreme points, one of the two (conflicting) objectives is favoured (or prioritised) in the array design. The corresponding array geometries are indicated, for each group of solutions, in the inlet images of Figure 1 (right). 

The results indicate that the different optimal arrangements of the 5-telescope array favour the objectives differently, giving rise to a set of optimal solutions. These solutions represent the trade-off between the weights that the array designer wishes to attribute to the objectives in question. This family of diverse solutions, that are the result of a single run of the algorithm, are one of the great strengths of the evolutionary multi-objective approach used here. Collectively, they present the decision maker with all the possible, optimised geometries that can be chosen from in building the best array. The final choice of which array geometry to adopt is a decision of the array designer, according to its specific scientific goals. But, thanks to the family of solutions provided by the algorithm, this will be a well-informed decision, taking into consideration a complete mapping of the parameter space of the problem in question. 

The groups of array geometries of interest can then be used, for example, as input to more detailed investigations with a complete Monte Carlo simulation. In this way, the evolutionary algorithm serves as an auxiliary tool to the experimental design, mapping the parameter space and directing the simulations to save computing time and help identifying potentially interesting, but non-intuitive solutions, that would hardly be considered otherwise.

The examples presented, with only a few telescopes and two objective functions, are to illustrate the technique. Its potential becomes much more evident, and the family of solutions, interesting, when simultaneously considering more optimisation objectives and a greater number of telescopes. The algorithm also  has the potential to: deal with different telescope types simultaneously in the optimisation; physical constraints restricting telescope positioning on the ground; and to treat the number of telescopes as a free variable of the problem.


\begin{thebibliography}{99}
\bibitem{Agrawal1995} Agrawal, R.B., Deb, K. and Agrawal, R. Complex Systems, v. 9 (2), p. 115, 1995.
\bibitem{Aharonian1997} Aharonian, F., Hofmann, W., Konopelko, A. and V\"{o}lk, H. Astroparticle Physics, v. 6, n.3, p.343, 1997.
\bibitem{Aleksic2016} Aleksic, J. et al.  Astroparticle Physics, v. 72, p. 76, 2016.
\bibitem{Bernlohr2013} Bernl\"{o}hr, K. Astroparticle Physics, v. 43, p. 171, 2013.
\bibitem{Coello2007} Coello, C.A.C., et al. Evolutionary algorithms for solving multi-objective problems, Springer, 2007.
\bibitem{Cohanin2004} Cohanin, B.E., Hewitt, J.N. and de Weck, O.L. ApJ SS, v. 154, p. 705, 2004.
\bibitem{Deb2001} Deb, K. Multi-objective optimisation using eviolutionary algorithms, John Wiley \& Sons, 2001.
\bibitem{Deb2014} Deb, K, Jain, H. IEEE Trans. Evolutionary Computation, v.18 (4), p. 577, 2014.
\bibitem{Heck1998} Heck, K. et al. CORSIKA: A Monte Carlo code to simulate extensive air showers. FZKA 6019. Technical Report, 1998.
\bibitem{Hillas1985} Hillas, M. Cherenkov light images of EAS produced by primary gamma, in: Proceedings of International Cosmic Ray Conference, v. 3, 1985.
\bibitem{Hillas2013} Hillas, M. Astroparticle Physics, v. 43, p. 19, 2013.
\bibitem{Ibrahim2016} Ibrahim, A., Rahnamayan, S., Martin, M.V., and Deb, K. EliteNSGC-III: An improved evolutionary many-objective optimisation algorithm. IEEE Congress on Evolutionary Computation, p. 973, 2016.
\bibitem{Konopelko1999} Konopelko, A. et al. Astroparticle Physics, v.10 (4), p.275, 1999.
\bibitem{Maier2017} Maier, G., Arrabito, L., Bernl\"{o}hr, K. et al. Performance of the Cherenkov Telescope Array, Proc. 35th ICRC (Busan, Korea), e-print arXiv:1709.03483.
\bibitem{Weekes1989} Weekes, T.C. et al. The Astrophysical Journal, v. 342, p. 379, 1989.

\end{thebibliography}
\end{document}